# PLACEMENT OF ENERGY AWARE WIRELESS MESH NODES FOR E-LEARNING IN GREEN CAMPUSES

G.Merlin Sheeba[1], Alamelu Nachiappan[2,] P.H.Pavan Kumar[3], Prateek [3]

[1, 3] Department of Electronics and Telecommunication Engineering, Sathyabama University, Chennai, Tamilnadu, India
[2] Departments of EEE, Pondicherry Engineering College, Puducherry

### ABSTRACT

*Energy efficiency solutions are more vital for Green Mesh Network (GMN) campuses. Today students are benefited using these e-learning methodologies. Renewable energies such as solar, wind, hydro has tremendous applications on energy efficient wireless networks for sustaining the ever growing traffic demands. One of the major issues in designing a GMN is minimizing the number of deployed mesh routers and gateways and satisfying the sustainable QOS based energy constraints. During low traffic periods the mesh routers are switched to power save or sleep mode. In this paper we have mathematically formulated a single objective function with multi constraints to optimize the energy. The objective is to place minimum number of Mesh routers and gateways in a set of candidate location. The mesh nodes are powered using the solar energy to meet the traffic demands. Two global optimisation algorithms are compared in this paper to optimize the energy sustainability, to guarantee seamless connectivity.*

## 1. INTRODUCTION

Wireless Mesh networks (WMN) shows a paradigm shift in development of wireless services for versatile users in person, enterprises, university campuses, in urban and rural scenarios. Green WMN has received a tremendous attraction in recent years. Renewable energies are playing a major role in green networking solutions. The widely deployed low cost WMN faces many challenging issues. Reducing unnecessary wastage of energy is one of the major concerns of today's economic world. To solve this problem it is essential to sustain the energy and minimize the energy consumption. The solution for an energy efficient Information Communication Technology (ICT) is to consider the equipments overall performance from manufacturing to its. endcycle and recycling [1].The network devices connected to the internet almost consume 70% of global telecommunication which will increase more in the coming decades[2].
The node placement problem is studied in recent years by formulating them as an optimization problem with different objectives and constraints. The objectives discussed in the literature so far are minimizing the number of mesh routers [3], [4], [5], maximizing the network throughput and connectivity [6],[7],[8], minimizing the deployment cost[9],minimizing energy consumption[10] etc.

The WMN cloud consists of wireless mesh routers, mesh clients, mesh access points and gateways otherwise referred as mesh points, mesh point portals [11].University campuses are deploying the cost effective WMN for its seam less connectivity. Mesh routers have minimum mobility which form the backbone of the WMN.The integration of WMN with Internet and other wireless networks such as IEEE 802.11,802.15,802.16 etc. can be done by gateway and bridge functions. Mesh clients can be mobile or fixed and they can form as a separate client network. The WMN infrastructure network devices are always active but at low traffic periods the energy consumption is same as the busy hours. Hence an energy aware design is necessary especially for university campuses. The network will be busy only during working hours at day time in campuses compared to night. Many methodologies are blooming for education in rural areas such as e-learning which utilizes the WMN techniques for connectivity.

## 2. RELATED WORKS

Renewable energy sources such as solar, wind, hydro can sustain the ever growing traffic demands. Zhongming Zheng et.al. [12] have formulated a Access Point (AP) placement optimization problem. The objective of the problem is to optimally place the minimum number of APs and meeting the QOS requirements through the harvested energy resources also. Based on the charging capabilities of the APs and users demands a rate adaptation and joint power control scheme is followed. An optimal network performance is achieved by using a polynomial based time complex heuristic algorithm. The proposed heuristic algorithm has shown better performance compared with the exhaustive search method. Sarra Mamechaoui et.al.[13] have discussed about the power management in WMN for developing Green environment which is considerably attracting the world nowadays. During low traffic periods, the unused Mesh Routers (MRs) are brought to sleep mode without affecting any available traffic paths. A Mixed Integer Linear Programming Model (MILP) is presented with the objective to minimize the number of mesh routers and thereby reduce the energy consumption.

The rechargeable router placement problem presented by Xiaoli Huan et. al.[14] is an optimization problem with the objective to minimize the number of deployed routers and satisfying the QOS constraints such as demand of users, energy consumption, network failure rates and traffic fairness. The authors have discussed about two cell association algorithms such as nearest cell association and proportional fairness algorithms to assign the users to the suitable routers. The proportional fairness algorithm tries to show a balance between the network performance and fairness. An extensive analysis is performed with association algorithm based router placement methods such as exhaustive search, greedy search based and Simulated Annealing (SA).

## 3. E-LEARNING

E-learning has the greatest technological potential to spread learning E-learning refers to learning that is delivered or enabled through electronic technology. With the internet boom, the benefits of e-learning are now accessible to the masses. Over the years, our dependency over internet services has increased exponentially. This has in turn, lead to the demand for uninterrupted, secure and foolproof network connectivity. In order to provide steady network access, a whole new architecture should be provided instead of a conventional, or rather traditional one. WMN technology has caused a paradigm shift in providing high bandwidth network coverage without compromising on efficiency. It distributes high speed Wi-Fi coverage through mesh access points. As a result, all the area in that particular village will receive 100% Wi-Fi coverage. The e-learning experience will no longer be limited by the length of cables. The traditional Wi-Fi network installed will provide an alternative solution to extend the coverage area. Integration of WiFi networks are designed and deployed for campuses which outperforms the traditional wired networks [11],[15].

International Journal on Cybernetics & Informatics (IJCI) Vol. 4, No. 2, April 2015

## 3.1. University Campus

The campus model of Sathyabama University is shown in figure 1.A hypothetical placement of routers and gateways are shown in the CAD model. The drawing made is as per the manual measured data collected from the campus for the research work. This work was motivated by the survey done in the campus regarding the WiFi connectivity [11].

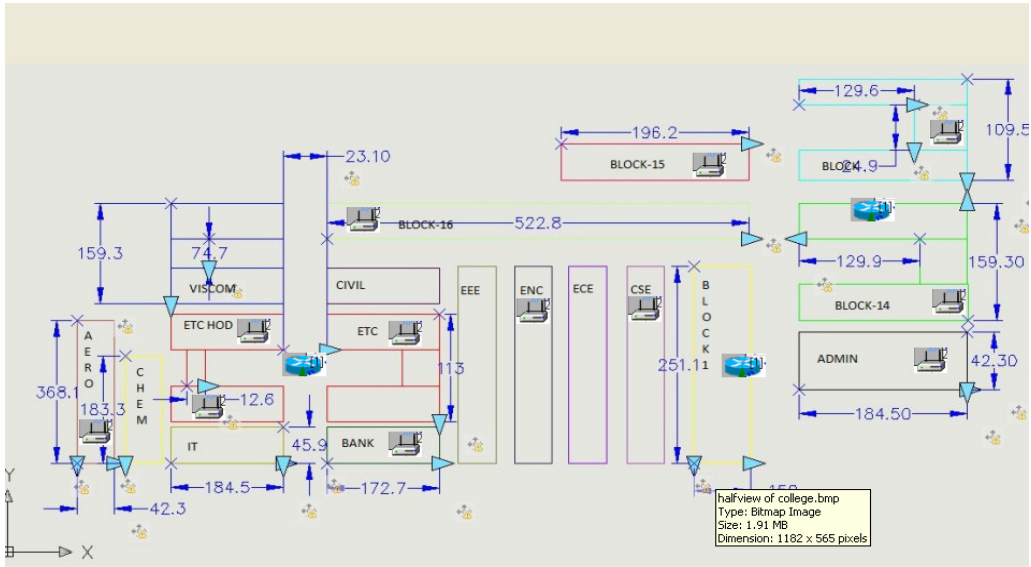

Figure 1. Hypothetical WMN nodes deployment in Sathyabama University

## 3.2. System Model

The campus mesh nodes in the path ways near the classroom blocks are powered with solar energy. Here the nodes are unevenly distributed in an irregular grid. The mesh clients get access through the mesh routers which are recharged using the solar energy. In figure 2 the system of sustainable energy calculation model flow is provided. The failure rate of each renewable energy sourced router is calculated with respect to threshold assigned initially for each node.

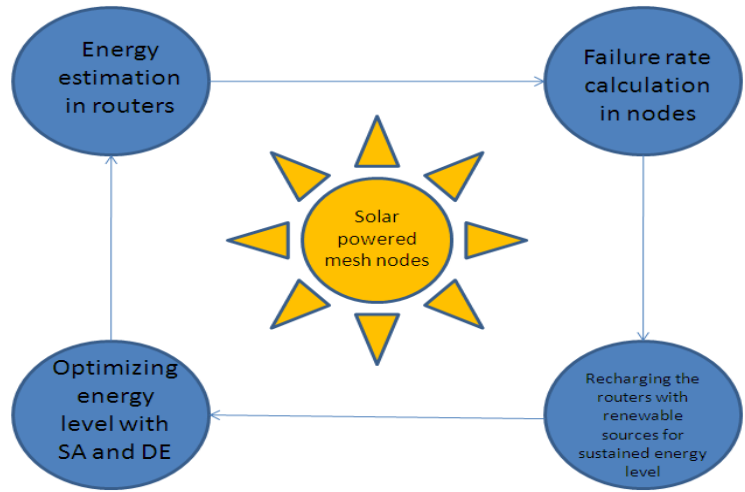

Figure 2.System of Suistanable energy Evaluation model flow

### 3.2. Mathematical Model Formulation

Our aim is to minimize the number of mesh nodes and thereby optimize the energy consumption throughout the working hours as well as in the non-working hours. The access networks have direct impact from the user end generated traffic profiles. When designing renewable energy sourced routers it is indeed a major design criterion to minimize the number of routers and satisfy the QOS constraints of the designed network.

Let S be the set of candidate locations where the mesh routers must be placed. S = {1………j}. The mesh routers act as gateway when that location has more traffic demand. The network can be considered as a connectivity graph G (V $\cup$ MR ,E). V represents the number of users and MR the mesh nodes and E denote the set of communication links between the nodes. $v \in V, mr \in MR$ Where $v$ is served by the $mr$

### 3.3. Problem Formulation

To place minimum number of mesh routers and gateways in the set of candidate locations. The nodes

$$Minimize \sum_{j \in S}(mr_j + g_j) \quad \text{(1)}$$

Where $mr_j$ is the $j^{th}$ is the mesh router, $g_j$ is the $j^{th}$ gateway
Subject to

**(i) Traffic Flow constraint**

Let $\ell_{ij}$ : is a binary variable denoting the total flow link between $v$ and mr

$$\ell_{ij} \in (0,1) \; v \in V, mr \in MR \quad \text{(2)}$$

**(ii) Energy consumption and sustainability constraint**

$$(e_{mrj} + e_{gj}) \leq E_{mrj} + E_{gj} \quad \text{(3)} \; \forall mr, g \in MR$$

Where $e_{mrj}$ the energy is consumed by the mesh routers and $e_{gj}$ is the energy consumed by the gateways for uplink and downlink. $E_{mrj}$ and $E_{gj}$ is the harvested energy for routers and gateways.

$$(e_{mrj} + e_{gj}) = P' \times T \quad \text{(4)}$$

Where $P'$ the average discharging rate of mesh router and gateway and T is the time period.

$$E_{mrj} + E_{gj} = P^{\oplus} \times T \quad \text{(5)}$$

Where $P^{\oplus}$ is the charging rate of the mesh router and gateway using renewable source.

**(iii) Failure rate(FR)**

$$FR \leq f_{th} \quad \text{(6)}$$

The constraint (6) ensures that the network failure rate must not be more than the predefined threshold decided by the network designer to maintain the coverage.

The FR calculation is incurred from [14] and it is given as:

$$\text{Failure rate (FR)} = \frac{\sum_{l \in T_s} \sum_{i \in V} (1 - \sum_{j \in S} a_{ij}(l))}{|V| * |T_s|} \quad \ldots\ldots\ldots\ldots(7)$$

Where, $\{a_{ij}(l) = 1$ if Mesh client(MC) is associated with the jth router
  $0$ otherwise

|V| is the total number of MC,
$T_s$ is the total number of time slots

## 4. NODE PLACEMENT METHODS

### 4.1. Greedy Placement Method

In greedy based placement method the routers are located one by one in the candidate locations. The failure rate is calculated in each stage[14]. The routers are located repeatedly and positioned until there is a minimum failure rate.

### 4.2 Exhaustive search based placement

For deployment of N routers, the failure rate can be computed with the a cell association algorithm [14].Among the iterations, the less number of routers with failure rate less than the assigned threshold is finalized as the optimal solution.

### 4.3 Simulated Annealing based placement

The method is modelled with the background of thermodynamics applications. The physical process involves the heating of material and then slowly cooling the material to minimize the energy of the system. After each iteration a new point or value is reached, the distance between the new and the current point is searched based on a probability. The new points are accepted which minimizes the objective and also the points which raise the objective is also included to avoid local minima.

```
Pseudo Code

Intialize
Begin
M=m0
For i=0 to mmax (or m min)
S ← m / m max
mnew ← neighbours(m / m max)
    if
    probability;
    P(E(m),E(mnew),S) > random(0,1)

    m ← mnew
    else
        output ← m
    end

end
```

Here in this paper the maximum no. routers initialized are 36. A 6x6 grid area deployment is planned with one mesh router in each cell. If the traffic demand is high in a particular cell, the router will act as a gateway. Then one by one the routers are removed until the FR exceeds the threshold value. In all the candidate locations S the routers are placed and the initial failure rate is calculated as $FR_0$. Then we decrease the number of routers by one and then a new set of placement is done and again the $FR_1$ is calculated. Then SA probabilistically decides to stay in which placement new or old. The selection process repeats until the satisfied amount of iterations.

### 4.4 Differential Evolution (DE) based placement

The DE algorithm is a global optimization technique [16], belonging to the general class of evolutionary algorithms. In the DE algorithm, there are conventionally three operators (i.e., mutation, crossover and selection) and three parameters, i.e., the population size NP, the scale factor F and the crossover probability CR. Mutation plays a key role in the performance of the DE algorithm and there are several variants of mutation.

Select the size of the population N (it must be at least 4)

• The control parameter vectors have the form:

$$xG = [x1, x2............ xD] \qquad (8)$$

D is the no of parameters and G is the generation number.

**Initialization**
Define upper and lower limits for each parameter
Select randomly the initial parameter values uniformly on the intervals
$[xLj , xUj]$

**Mutation:**

- All the parameter vectors undergo mutation, recombination and selection. Mutation expands the search space.
- For a given parameter vector xi randomly select three vectors xr1, xr2 and xr3
- Calculate the mutated vector

$$Vm = xr1 + S (xr2 − xr3) \qquad (9)$$

- The mutation factor S is a constant from [0, 2].
- Vm is called the donor vector.

**Recombination/Cross Over:**

- Crossover produces successful solutions from the previous generation.
- The trial vector YT is developed from the elements of the target vector xi and the elements of the donor vector Vm.
- Elements of the donor vector Vm enter the trial vector YT with probability CR

- $YT = \begin{cases} Vm & \text{if } \eta_j \leq CR \text{ or } j = I_{rand} \\ x_i & \text{if } \eta_j > CR \text{ or } j \neq I_{rand} \end{cases}$ (10)

$i = 1, 2. . . N; j = 1, 2. . . D$
$\eta_j$........ any random number between 0 and 1
Irand is a random integer from [1, 2... D]

**Selection:**

- The vector xi is compared with the trial vector YT and according to the objective function the maximum or minimum value will be in the next generation.

$$x_{i+1} = \begin{cases} Y_T & \text{if } f(Y_T) \leq f(x_i) \\ X_i & \text{otherwise} \end{cases} \quad (11)$$

i=1,2,………………,N
- Mutation, recombination and selection procedures continue until some stopping criterion is achieved.

Pseudo Code

```
Begin
I=0
(Initialize Upper and lower limits of constraints)
Initialize the no. Of population size P
For I = 1 to generations do
For i= 1 to P do
Select randomly Ya ≠ Yb ≠ Yc
jrand = randint(1,D)
For j = 1 to D   do
    If (randj[0 1]) ≤ CR or j rand =) then
           Yj ,G+1 ← xYaj,G+S(xYb j,G – xYc j,G)
    Else
           Y i,G+1 = x j,G
    Endif
End For
    If f(Y i,G+1) ≤ f(Y i,G)  then
    X G+1 = Y G+1
     Else
    X G+1 = xG
    End If
End For
I=I+1
End For
End
```

## 5. RESULTS AND DISCUSSIONS

The network model using NS2 was created and the simulation results are observed using the optimization algorithms such as simulated annealing and Differential evolution. The scenario setup is based on a square field size of 1000mx1000m in which the mesh clients are uniformly distributed. All the routers have uniform charge. We repeat the simulation with different iterations and observe the failure rate with respect to the threshold as shown in figure 3. The results of both SA and DE are compared. It is observed that the DE gives less failure rate compared to SA.The simulation settings are shown in Table 1.

Table 1: Simulation Settings

| Parameters | Simulated Annealing | Differential Evolution |
|---|---|---|
| Placement of Nodes | random | random |
| Population Size | 100 | 100 |
| Cross over constant | Probabilistic selection | 0.5 |
| Scaling factor |  | 0.6 |

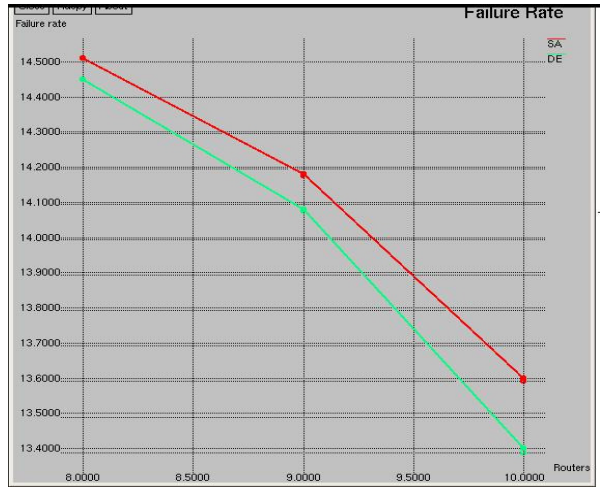

Figure 3. Failure rate calculation using SA and DE

The maximum charging capacity of each rechargeable router is taken as 100MW. To further observe the evaluation and feasibility of the algorithms used, the sustainable energy is observed in a time scale, shown in figure 4. The best values are plotted after each iteration. The DE algorithm converges faster compared to other global optimization technique SA. In SA after each placement scheme if the FR constraint is not satisfied the routers are decreased one by one and energy is monitored.

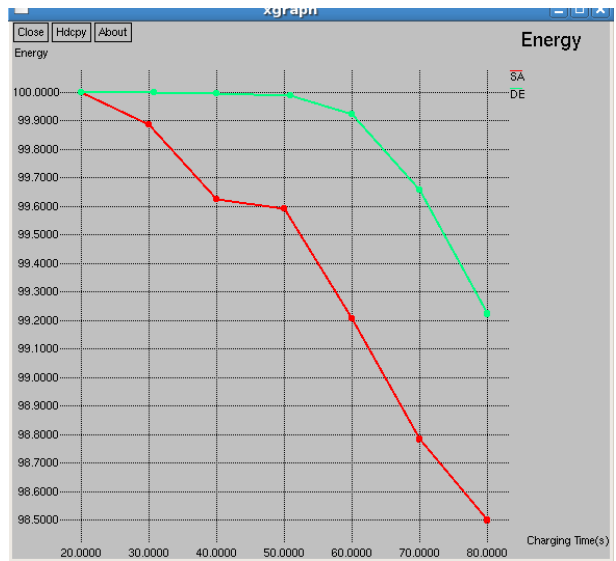

Figure 4. Energy sustained in Rechargable Routers using SA and DE

## 6. CONCLUSION AND FURTHER RESEARCH WORK

In this paper we have studied about mesh router placement problem for e-learning in campuses. We have formulated an optimization problem to minimize the number of gateways and mesh routers with energy sustainable constraint. The key objective of the proposed formulation is to plan and design an energy efficient GMN. The two global optimisation techniques SA and DE is studied and used for evaluation of rechargeable routers to reach an optimal failure rate. As university campuses are currently deployed with energy aware nodes more research area is still open for the networking community. From the simulations it is obvious that DE outperforms than SA, where both are named for their global optimum results. The work can be further extended by using hybrid DE algorithm and multiobjective optimizations.

## ACKNOWLEDGEMENTS

The author would like to thank the management Sathyabama University especially to the honorable chancellor Dr. Jeppiaar, Directors Dr. Marie Johnson and Dr. Mariazeena Johnson, for providing the necessary facilities for carrying out this research.